\documentclass[prl,superscriptaddress,amsmath,amssymb,twocolumn]{revtex4}

\usepackage{bm}
\usepackage{amsfonts}
\usepackage{amsmath}
\usepackage{graphicx}
\usepackage{dcolumn}

\begin{document}

\title{Bound entanglement in the XY model}

\author{D. Patan\`e}
\affiliation{MATIS-INFM $\&$ Dipartimento di Metodologie Fisiche e
Chimiche (DMFCI), Universit\`a di Catania, viale A. Doria 6, 95125
Catania, Italy}
\author{Rosario Fazio}
\affiliation{International School for Advanced Studies (SISSA), via Beirut 2-4,
 34014 Trieste Italy}
\affiliation{NEST-INFN $\&$ Scuola Normale Superiore Piazza dei Cavalieri 7, I-56126 Pisa, Italy  }
\author{L. Amico}
\affiliation{MATIS-INFM $\&$ Dipartimento di Metodologie Fisiche e
Chimiche (DMFCI), Universit\`a di Catania, viale A. Doria 6, 95125
Catania, Italy}

\begin{abstract}
We study the multi-spin entanglement  for the 1D anisotropic
XY model concentrating on the simplest case of three-spin entanglement. As compared to the pairwise
entanglement, three-party quantum correlations have a longer range and
they are more robust on increasing the temperature. We find  regions
of the phase diagram of the system where bound entanglement occurs, both at
zero and finite temperature. Bound entanglement in the ground state can be
obtained by tuning the magnetic field. Thermal bound entanglement emerges
naturally due to the effect of temperature on  the free ground state entanglement.

\end{abstract}


\maketitle

Entanglement is a resource in quantum information science~\cite{RESOURCE}.
More recently it has become clear that understanding the nature of quantum
correlations may also help in a deeper description of complex many-body
systems (see Ref.~\cite{REVIEW} for a recent review on this topic).
The work developed in these years in between the two areas of  quantum
statistical mechanics and quantum information has lead to several interesting results in
both disciplines. Among the different aspects investigated so far we mention
the study of entanglement close to a quantum phase transition~\cite{NATURE,OSBORNE02,
VIDAL}. Most of the work so far was
developed to bipartite entanglement with some notable exceptions.
On the other hand the entanglement monogamy property \cite{MONOGAMY,KKW}
constrains the entanglement sharing, and put an upper bound to the pairwise entanglement.

Several indications demonstrate that the multipartite entanglement
is indeed particularly important for the collective behavior of the
system. It was shown, for example, that  multipartite entanglement
is enhanced with respect to the two-particle entanglement near a
quantum critical point\cite{OSBORNE02,FUBINI,anfossi}. Multipartite
entanglement close to quantum  phase transitions was quantified by
the global-entanglement measure of Meyer and Wallach in
~\cite{oliveira} or the geometric measure of
entanglement~\cite{wei05}. The generalized entanglement measure
introduced in~\cite{barnum} was used to study a number of
critical~\cite{somma04} and disordered~\cite{montangero} spin
models.

A comprehensive classification of the type of multipartite entanglement in spin systems
has been recently given by~\cite{guehne05,TOTH}. Different bounds obtained to the
ground state energy were obtained for different types of n-particle quantum correlated states.
A violation of these bounds implies the presence of multipartite entanglement in the system.
As discussed in~\cite{facchi} the analysis of the average measures of
multipartite entanglement might not be sufficient and the analysis of the distribution of
block entanglement  for different partitions may give additional information.

In the present paper we  focus on the Multiparticle Entanglement
(ME) with the aim to shed light on  how entanglement is shared in  a
many body system. Specifically we analyse the type of entanglement
of  a subsystem made of few particles, tracing out the rest of the
system and we study the ME shared between them (this approach was
recently carried out also for the free electron
gas~\cite{VEDRAL-FERMIGAS,VERTESI}). By this approach, although we
cannot discuss the global ME properties of the whole system, we can
gain insight in the details of such few-particle ME. In particular
we consider the 1D XY model in transverse field and we study the
simplest multiparticle case  of a subsystem made of three arbitrary
spins of the chain.
 We analyze bipartite entanglement between a spin and the other two with respect to all possible bipartitions.
We demonstrate that ME  extends over a longer range than
two-particle entanglement. An important feature, emerging from our
analysis of  ME, is the existence of Bound Entanglement (BE)
\cite{HORO1,DUR-CIRAC} shared among  the spins of the chain. This
peculiar form of entanglement is characterized by the impossibility
of distilling it into a pure form.  It is  a weak form of
entanglement having features of  both quantum and classical
correlations. Bound entanglement has been a subject of intense
research in the last years since it was shown to be a useful
resource in the context of quantum
information~\cite{ENTA-REVIEW,HORO2,HORO3,BOUND1, MACCHIAVELLO}.

 Here we show that it emerges from the equilibrium properties of
spin chains. Indeed  for the ground state of quantum XY model
 spins  far apart enough may be in a bound entangled state, even in the
thermodynamical limit. At   $T\ne0$ we find that ME  is  more robust than
two-particle entanglement and increasing the temperature it always turns into a thermal
multiparticle bound entanglement.

The paper is organized as follow: in the following section we  review the  scenario of
three-qubit entanglement; we then briefly describe the  model studied and some  known
results concerning its entanglement properties.
In the last section we show our results  at $T=0$ and at finite temperature.

\section{Free and Bound Three-qubit Entanglement}
\label{section-entanglement} For  two-qubit systems,
Concurrence~\cite{WOOTTERS} is a  measure of entanglement that can
be easily calculated both for pure and mixed states and
 it can be related to the 'entanglement of formation'~\cite{SHUMI}.
Besides, any two-qubit entangled state can be converted into Bell states
with the process of distillation~\cite{SHUMI, NPTBOUND}.

For the   entanglement of a three-qubit system, a much more complex scenario emerges.
 The total entanglement shared between the three qubits
cannot be described just by  studying the qubit/qubit entanglement between all the pairs,
as measured f.i. by the Concurrence of the reduced 2-qubit density matrices.
 In fact in general the three qubits share a ME whose complete information is
unavoidably lost by tracing out a qubit.  The well known paradigmatic example are
the GHZ  states for which  each  couple of spins shares no entanglement,
nonetheless each spin is maximally entangled with the other two~\cite{KKW}.
A further qualitative difference that emerges for the three qubits entanglement
is the existence of bound entangled states.  The classification  'free' and 'bound'
entanglement can be drawn with respect to distillation properties of the entanglement~\cite{HORO1}.
Entanglement is bound (not-distillable) if no maximally entangled states
 between the parties of the system can be obtained with local operations and
classical communication (LOCC),  not even with an asymptotically
infinite supply of copies of the state~\cite{DUR-CIRAC}. Despite
thepragmatic definition of BE, the nature of its correlations is
peculiar. In fact BE is a very weak form of entanglement that has
both quantum and classical features.
 For instance  some examples of multipartite bound entangled state
violating Bell-type inequalities were found~\cite{DUR2}, but
recently it was shown that if we allow collective manipulations, and
postselection, no bound entangled state violates Bell-type
inequalities \cite{MASANES2}.

While for two-qubit system no BE states exist~\cite{NPTBOUND},
in the case of a three-qubit system the relative Hilbert space is large enough
to allow such structure to appear.
An interesting case in which such BE appears is related to the 'incomplete separability'
of the state (see~\cite{DUR-CIRAC}).
 This condition happens for example  when a state of a tripartite system A-B-C is separable
with respect to the partition A|BC and B|AC and non-separable with respect to C|AB.
The 'incomplete separability' is a sufficient condition for a state to have  BE
since the three qubits are entangled and
no maximally entangled state can be created between any of the parties by LOCC.
 For example, no entanglement can be  distilled  between C and A
because  no entanglement can be created with respect to the partition A|BC by LOCC.
 In the following sections,  the feature of incomplete separability will be exploited to detect
such a kind of BE in the system.

In order to analyze the free and bound entanglement between the qubits  we focus on the entanglement
 between a qubit and the other two with respect to all possible bipartitions.
To measure such   bipartite entanglement we used the Negativity~\cite{WERNER}:
\begin{equation}
\mathcal{N}_\mathcal{P}=\sum_i|\mu_i|,\label{neg}
\end{equation}
where $\mu_i$ are the negative eigenvalues of the partial transpose
 of the density matrix with respect to the bipartition $\mathcal{P}$.
Negativity gives the degree of violation of Peres condition and
 was proved to be  a  measure of  entanglement  \cite{WERNER,EISERT-PHD}, though
it can not detect Positive Partial Transpose (PPT) entangled states.

 We remark that
 the genuinely \emph{multipartite} entanglement (e.g. GHZ type of
 entanglement) is distinct
from generic multiparticle entanglement (e.g. a mixture of Bell
states shared between the parties)\cite{NOTE}.  In \cite{FAN} a
Negativity-based multipartite entanglement measure  was proposed,
but  it is not easy to handle for  generic mixed states since it
requires a convex roof minimization. For a multiparticle system,
Negativity is able to measure  the 'global' ME with respect to a
certain  bi-partition.
 For instance,  Negativity  can not distinguish between qubit-qubit entanglement and  ME.
For example, if  qubit A is entangled separately  with qubit B
(as measured by a non zero concurrence $C_{AB}$)
 then A will surely be entangled with B and C as a whole and thus $\mathcal{N}_{A|BC}\ne 0$.
On the other hand if {\it no qubit/qubit entanglement} is present in the system, then  a
{\it non-zero Negativity  detects multiparticle entanglement}.
Besides,  carrying such analysis  with respect to all possible bipartitions
can give information on how such ME is shared between the qubits.  It is precisely
this  method that we will employ to study
the ME entanglement in the system.

\section{The XY model}

Entanglement  has been studied in a variety of  spin models~\cite{REVIEW}.
A lot of attention has been devoted to the 1D quantum XY model. The Hamiltonian is defined as
\begin{equation}
H=J\sum_{i=1}^N (1+\gamma)S^x_i S^x_{i+1}+
(1-\gamma)S^y_i S^y_{i+1} - B S^z_i. \label{model}
\end{equation}
The spins $S^{\alpha}=\frac{1}{2}\sigma^{\alpha}$, $\alpha=x,y,z$
($\sigma^{\alpha}$ are Pauli matrices) experience an exchange
interaction with coupling $J>0$ and uniform magnetic field of strength
$B$. In the thermodynamical limit  $N\rightarrow \infty$
the model  (\ref{model}) has a quantum critical point at $h\doteq B/2J=h_c=1$, and for
$0<\gamma \le 1$ it belongs the quantum-Ising universality
class~\cite{MODELLO}.

Entanglement properties  of the model have been   studied,
particularly close to its phase transition \cite{REVIEW}. Attention
has been  focused in the entanglement between two spins or the
entanglement between  a block of spins and the rest of the lattice.
Spin-spin entanglement was studied by means of concurrence and it
was shown that a spin is directly entangled to its neighbor spins
and the range of spin-spin entanglement  $R$ (i.e. the maximal
distance at which two spins are still entangled) depends on  $h$ and
$\gamma$ \cite{NATURE,DIV}. In particular  for the ground state  $R$
diverges at the factorizing field~\cite{KURMANN}
$h_f=\sqrt{1-\gamma^2}$~\cite{DIV} while it is non universal and
short ranged at critical point~\cite{NATURE}. Block entropy was
studied in terms of Von Neumann entropy and it was shown that it
saturates as  the size of the block increases for a non critical
system, while it diverges logarithmically at the critical point with
a universal behavior ruled by its conformal
symmetry\cite{VIDAL,JIN-KOREPIN}.

We consider the system defined in Eq.(\ref{model})  in the thermodynamical limit
and we focus  our attention in a subsystem made of three spins
in different positions in the lattice (not necessarily a block).
We are interested in the entanglement shared between the three spins,
and  how such ME is distributed along the chain.
Thus, as sketched in the previous section, we compute the Negativity
between a spin and the other two with respect to all possible bipartitions of the three spins.
A similar analysis was done in~\cite{VEDRAL-FERMIGAS} for three fermions out of a Fermi gas.

\section{Three-spin Entanglement in the XY model}
\label{section-3entanglement}
Due to the complete  integrability of the  XY model,
 it  is possible to calculate the reduced density matrix of any number of spins
(see Appendix for the details).
The density matrix is a function of the parameters, $h$ and $\gamma$, the temperature $T$
and of the distances between the spins considered: $\alpha=j-i$ and $\beta=k-j$.
We first study the case $T=0$ focusing on the range of the three-spin entanglement and
then we consider the case of a block of three spins at finite temperature.

\subsection{Ground State Entanglement}

\begin{figure}[ht]\centering
\includegraphics[width=5cm]{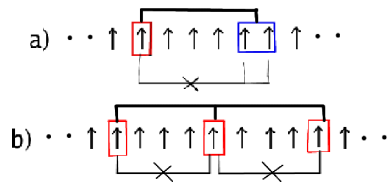}
\caption{ Configurations of spins described  in the text and whose entanglement properties
are  presented in Fig. \ref{NegT=0}.
 We fix the range of spin/spin entanglement $R\leq 3$ such that spins at a distance $d\geq4$
are not directly entangled.
 {\bf a)}  'Clustering' two spins increases the range of entanglement
(the scheme is symmetric also for spins on the left of the  marked one).
{\bf b)} Symmetric configuration of spins such that no two-particle entanglement is present,
 but still the spins share multiparticle entanglement.}
\label{clustering}
\end{figure}

\begin{figure}[ht]
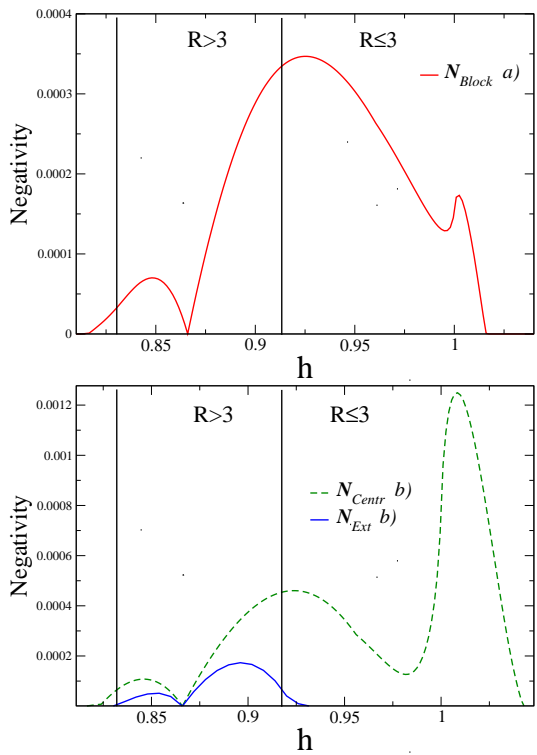
\centering
\includegraphics[width=7cm]{Nega.eps}
\includegraphics[width=7cm]{Negb.eps}
\caption{$T=0$ Negativities between one spin and the other two Vs magnetic field $h$ are shown
for both configurations of Fig \ref{clustering}.
 We consider $\gamma=0.5$. In this case outside the interval
 marked by the two solid vertical lines ( $0.83 \lesssim h\lesssim 0.91$)
the range of spin-spin entanglement is $R\leq 3$
(for values of $h$ inside this interval  $R$ grows
due to the its divergence at factorizing field $h_f=\sqrt{1-\gamma^2}\simeq 0.86$ \cite{DIV}).
For configuration $a)$ (upper panel), $\mathcal{N}_{Block}$  signals the spin/block entanglement for a distance $d=4$.
For values of $h$ outside  the vertical lines, the Negativity signals genuine spin/block entanglement.
For configuration $b) $ (lower panel), both  $\mathcal{N}_{Ext}$ (solid line) and  $\mathcal{N}_{Centr}$ (dashed line) are plotted.
For values of $h$  outside the vertical lines the three spins share no spin/spin entanglement,
hence  for  non zero $\mathcal{N}_{Ext}$ and $\mathcal{N}_{Centr}$
 free multiparticle entanglement is present.
 The latter turns in to bound entanglement for values of  $h$ such that only
$\mathcal{N}_{Centr}\ne 0$ and  $\mathcal{N}_{Ext}=0$ (both on the left and on the right of the solid lines).  }
\label{NegT=0}
\end{figure}

We first focus on the range of three-particle entanglement in the ground state and
check whether  spins distant enough such that
they do not share spin/spin entanglement are nonetheless globally entangled.
 With this aim,  we consider a spin and a block  of two spins more distant than
the spin/spin entanglement range (see Fig.\ref{clustering} a))
and we  study the  spin/block entanglement as measured by the Negativity.
 We  find that  the block of two spins  may be  entangled with the  external spin,
despite the latter  is not entangled directly with any of the two spins separately
(see Fig. \ref{NegT=0} upper panel).
Hence the range of such spin/block entanglement  may extend further than
the spin/spin entanglement range.

An interesting case is when there is no  two-particle entanglement between
of any the three spins, but still  they share ME.
Specifically, we consider the symmetric configuration of spins drawn in Fig.\ref{clustering}b.
As shown in Fig.\ref{NegT=0} (lower panel), the Negativity between the central spin and the other two $\mathcal{N}_{Centr}$
  and between an external spin and the other two  $\mathcal{N}_{Ext}$ may be   non-zero,
 despite the three spins do not share spin-spin entanglement.

We shall see that spins in the configuration  of Fig. \ref{clustering} b) can share bound entanglement.
To prove it, the idea is to resort  the "incomplete separability" condition described in the first section.
In fact from  Fig. \ref{NegT=0} we see that  $\mathcal{N}_{Ext}$  may be zero even if  $\mathcal{N}_{Centr}$ is non-zero.
 Thus in such case the density matrix of the spins is  PPT for the two symmetric bipartitions of one external spin vs the other two
($\uparrow|\uparrow \uparrow$ and $\uparrow\uparrow |\uparrow$) and
 Negative Partial Transpose  (NPT) for the partition of the central spin vs the other two.
We remark that PPT does not ensure the separability of the two partitions,
nevertheless the state is bound entangled.
In fact if we could be able to distill a maximally entangled state between two spins
  then one of two previous PPT partitions would be NPT
and this cannot occur since PPT is invariant under LOCC \cite{HORO1,WERNER}.

The scenario described above  quantitatively varies if different
values of $\gamma$ are considered. Since the range of spin-spin
entanglement depends on $\gamma$~\cite{DIV} the distance between the
spins at which ME exists  and its range are also $\gamma$-dependent.
 For both the configurations we studied we found that ME  is
short ranged.
 However,  we notice that such range diverges  at the factorizing
field $h_f$ (analogously to what occurs for  spin/spin entanglement
\cite{DIV}). Remarkably, this   holds also for  BE.
  In fact we observe that the range of $\mathcal{N}_{Centr}$ is always greater than  the range of $\mathcal{N}_{Ext}$.
   Hence approaching $h_f $  there are always configurations of spin far
 apart enough to share BE and  for $h\rightarrow h_f$ its range
 diverges.
\\

In summary if the spins are far enough to loose all spin-spin entanglement,
some ME may be  still present. The nature of the entanglement   may change from free
to  bound entanglement by tuning the magnetic field $h$.

\subsection{Thermal Bound Entanglement}

\begin{figure}[ht]\centering
\includegraphics[width=8cm]{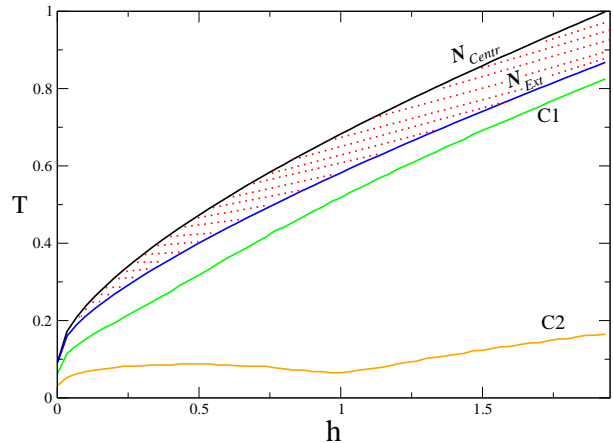}
\caption{
 Entanglement shared in a block of the three adjacent spins.
We consider $\gamma=1$. In this case only nearest neighbor and next nearest neighbor spins are entangled
(hence $R=2$) at $T=0$ \cite{NATURE,OSBORNE02}.
The lines in the $T-h$ plane indicate the temperatures at  which the corresponding type of  entanglement disappears.
In the marked region  $T_{\mathcal{N}_{Ext}}<T<T_{\mathcal{N}_{Centr}}$ BE  is  present.}
\label{thermal}
\end{figure}

Quantum states must be  'mixed enough'  to be bound entangled. In
fact in a geometrical picture they may be  located in a region
separating free entangled states from the separable
ones~\cite{GEOMETRICAL-BOUND}.  In our case, a source mixing is the
trace over the other spins of the chain: as pointed out in the
previous section, such loss of information may be enough to induce
bound entanglement between the three spins. However  if the spins
are near enough the reduced entanglement is free. Thus in this
situation it is intriguing to investigate whether the effect of the
thermal mixing can drive the $T=0$  free entanglement to BE. To
answer this question we consider a block of three adjacent spins and
study the entanglement between them as a function of $T$. We see
(Fig. \ref{thermal}) that  the entanglement between the two external
spins is the most fragile increasing $T$, dying first at a certain
$T_{C2}$; then at a higher temperature $T_{C1}$ the central spin
looses its entanglement  with  each of the external ones. Thus for
$T>T_{C1}$ there is no two-particle entanglement between the spins,
but still some ME is present, as detected by the Negativity
$\mathcal{N}_{Ext}$ and $\mathcal{N}_{Centr}$. Eventually  the
Negativity vanishes at  $T_{\mathcal{N}_{Ext}}$ and
$T_{\mathcal{N}_{Centr}}$. In the region
$T_{\mathcal{N}_{Ext}}<T<T_{\mathcal{N}_{Centr}}$ the condition of
'incomplete separability'  occurs: the density matrix is PPT with
respect to the external bipartitions and NPT with respect to the
central one and thus the spins share BE.

We want to remark here that the failure of Negativity to detect some entangled states
does not invalidate the scheme described above.
 In fact, when all the Negativities are eventually zero, the three-spin state may be  still entangled,
but such entanglement would be surely bound (since the state has positive partial transpose).
This means that the region where bound entanglement is present
might be wider than the one marked in Fig. \ref{thermal},
but, any case, such thermal bound entanglement will always separate
the high temperature separable region from the low temperature free entangled one.

This behavior shown for the Ising model is also found  for the
entire class of Hamiltonians (\ref{model}). In fact although the
temperature at which the different types of entanglement  are
suppressed is $\gamma$ dependent the qualitative behavior of the
phenomenon is a general feature of the model (\ref{model}).

\section{Conclusions}

We studied the entanglement shared between three spins in an infinite  chain
described by the anisotropic XY model in a transverse field.
We analyzed the Negativity between groups of spins as in Fig.\ref{clustering},
demonstrating  that  the corresponding block/block entanglement extends
 along the chain over a longer range (Fig. \ref{NegT=0}) compared to the spin/spin case.
Such type of entanglement persists at  higher temperature (see Fig. \ref{thermal}).

 In a recent paper, thermal BE  was found for system made of few
spins~\cite{TOTH07}. Here we  proved the existence of bound
entanglement shared between the particles, even in a macroscopic
system.  It occurs naturally in certain region of the phase diagram
for the 'last' entangled states before the complete separability is
reached. In this sense the BE bridges between quantum and classical
correlations. Being the BE  a form of demoted entanglement, we found
that it appears when quantum correlations get weaker. At zero
temperature bound entanglement appears  (see Fig. \ref{NegT=0}) when
the spins are sufficiently distant each others  and as in the case
of the spin/spin entanglement, it can be arbitrary long ranged near
the factorizing field. At $T\ne0$ BE is present when the system is
driven toward a completely mixed state as the temperature is
raised(Fig. \ref{thermal}). It would be an interesting to further
study  bound entanglement in this context (f.i. focusing on  other
kinds of bound entanglement \cite{UPB}).

A more refined  classification of ME
could be done following the scheme of Ref.\cite{ACIN},
discerning different regions in the phase diagram in terms of different entanglement classes
(GHZ or W). In Ref. \cite{ACIN} certain witness operators able to distinguish
some $W$- and $GHZ$- mixed states  were discussed.
We point out, however, that in our case   such witness are not able to detect tripartite entanglement,
not even if suitably generalized as~\cite{VERTESI}.
Instead, the three tangle\cite{KKW}, that in principle could identify the GHZ class,
is difficult to handle for generic mixed state and requires a hard numerical effort
(some progress is achieved  for low rank density matrices\cite{JENS}).

It is plausible that increasing the size of the subsystem  considered
 will increase the range of the ME.
For instance, the range of  spin/block entanglement (see configuration of Fig. \ref{clustering} a)
  will increase if we consider a larger block.
Hence, a  single spin  can be  entangled  with more distant partners,
if one allow to cluster them into a large enough block.
It would be  intriguing  to study how   spin/block entanglement and,
 in general, block/block entanglement between subsystems,  scale increasing the size of blocks.
especially exploring the  connection with quantum criticality.
We remark that such analysis would be different
with respect to the well known block entropy setting\cite{VIDAL,BLOCK},
since in that case one is interested in the block/rest-of-the-system entanglement.
Along this line, some results for a chain of coupled harmonic oscillators were obtained\cite{WERNER-CHAIN}.

Finally,  we  speculate  about the  effect of temperature
for the entanglement of a larger size block of spins.
The behavior depicted in Fig. \ref{thermal} suggests that
entanglement  between sub-blocks of few spins will be suppressed by increasing the temperature, before
of the entanglement between larger sub-blocks.
Namely, temperature will suppress first spin/spin entanglement,
then spin/2-spinblock entanglement,.., n-spinblock/m-spinblock entanglement, and so on.
Thus,  at high enough temperature, particles of the system
will eventually loose all entanglement passing through successive steps
at which temperature suppresses  entanglement with respect to  a
'microscopic' to 'macroscopic' hierarchy.

\acknowledgments

We thank  J. Siewert for significant help. We acknowledge fruitful discussions with
A. Osterloh, F. Plastina. The work has been supported by PRIN-MIUR  and it has been
performed within the
``Quantum Information'' research program of Centro di Ricerca Matematica
``Ennio De Giorgi'' of Scuola Normale Superiore.

\section*{Appendix}

It is
convenient to express $\rho_{ijk}$ in terms of three-point correlation functions
that   can be obtained explicitly following the method used for the two-point
correlation functions in \cite{MODELLO}
\begin{equation}
\rho_{ijk}=
\sum_{pqr} \langle \sigma^p_i\sigma^q_j\sigma^r_k\rangle\sigma^p\otimes\sigma^q\otimes\sigma^r \;\;.
\label{expansion}
\end{equation}
In the previous equation  $p$,$q$,$r$$=0,x,y,z$ ($\sigma^0=\textbf{1}$).

Three-spin reduced density matrix is obtained from Eq.(\ref{expansion}).
Due to the parity symmetry of the Hamiltonian \cite{VIDAL2}, for the non-broken symmetry case
 some of the correlators are identically zero and the $8\times 8$ matrix reads:
\begin{displaymath}
\rho_{ijk}=\frac{1}{8}\left (
\begin{array}{cc}
\rho_I&\rho_{III}\\
    \rho_{III}&\rho_{II}
\end{array}\right ),
\end{displaymath}
where
 \begin{eqnarray*}
\rho_I&=& \nonumber \\
&&\hskip-35pt \left (
\begin{array}{cccc}
1+\mathcal{A}+\mathcal{B}^{++} & 0 & 0 &\mathcal{D}^-_\beta+\mathcal{E}^-_{\beta\alpha}\\
0 & 1+\mathcal{C}+\mathcal{B}^{--} & \mathcal{D}^+_\beta+\mathcal{E}^+_{\beta\alpha} & 0\\
0 & \mathcal{D}^+_\beta+\mathcal{E}^+_{\beta\alpha} & 1+\mathcal{C}-\mathcal{B}^{+-} & 0\\
\mathcal{D}^-_\beta+\mathcal{E}^-_{\beta\alpha} & 0 & 0 & 1-\mathcal{C}-\mathcal{B}^{-+}\\
\end{array}\right )
\end{eqnarray*}
 \begin{eqnarray*}
\rho_{II}&=& \nonumber \\
&&\hskip-35pt \left (
\begin{array}{cccc}
1+\mathcal{C}-\mathcal{B}^{-+} & 0 & 0 &\mathcal{D}^-_\beta-\mathcal{E}^-_{\beta\alpha}\\
0 & 1-\mathcal{C}-\mathcal{B}^{+-} & \mathcal{D}^-_\beta-\mathcal{E}^+_{\beta\alpha} & 0\\
0 & \mathcal{D}^-_\beta-\mathcal{E}^+_{\beta\alpha} & 1-\mathcal{C}+\mathcal{B}^{--} & 0\\
\mathcal{D}^-_\beta-\mathcal{E}^-_{\beta\alpha} & 0 & 0 & 1-\mathcal{A}+\mathcal{B}^{++}\\
\end{array}\right )
\end{eqnarray*}
\begin{eqnarray*}
\rho_{III}&=& \nonumber \\
&&\hskip-20pt \left (
\begin{array}{cccc}
0 &\mathcal{D}^+_\gamma+\mathcal{F}^+_{\alpha\beta} &\mathcal{D}^+_\alpha+\mathcal{F}^+_{\alpha\beta} &0\\
\mathcal{D}^-_\gamma+\mathcal{F}^-_{\alpha\beta} & 0 & 0 & \mathcal{D}^+_\alpha-\mathcal{F}^+_{\alpha\beta}\\
\mathcal{D}^-_\alpha+\mathcal{F}^-_{\alpha\beta} &0  & 0& \mathcal{D}^+_\gamma+\mathcal{F}^+_{\alpha\beta}\\
0 & \mathcal{D}^-_\alpha-\mathcal{F}^-_{\alpha\beta} & \mathcal{D}^-_\gamma-\mathcal{F}^-_{\alpha\beta} & 0\\
\end{array} \right ).
\end{eqnarray*}
The entries of the matrices above are linear combination of the correlation functions:
\begin{eqnarray*}
\mathcal{A} & = & 3Z+ZZZ\\
\mathcal{B}^{\pm\pm} & = & ZZ_\alpha\pm ZZ_\beta\pm ZZ_\gamma\\
\mathcal{C} & = & Z-ZZZ\\
\mathcal{D} ^{\pm}_\alpha& = & XX_\alpha\pm YY_\alpha\\
\mathcal{E} ^{\pm}_{\alpha\beta}& = & XXZ_{\alpha\beta}\pm YYZ_{\alpha\beta}\\
\mathcal{F} ^{\pm}_{\alpha\beta}& = & XZX_{\alpha\beta}\pm YZY_{\alpha\beta}
\end{eqnarray*}
where f.i. $YYZ_{\alpha\beta}= \langle \sigma^y_i\sigma^y_j\sigma^z_k\rangle$ with $\alpha=j-i$ and $\beta=k-j$
(the correlators depend only on the relative distance between the spins
 because of the translational symmetry of the system).

Such three-point correlation functions can be calculated following the method used in \cite{MODELLO} for
the two-point ones.
In short, after the Jordan-Wigner transformation which maps spin  into spinless fermions
\begin{eqnarray*}
\sigma_l^x&=&A_l \prod_{s=1}^{l-1} A_s B_s \nonumber \\
\sigma^y&=&-B_l \prod_{s=1}^{l-1} A_s B_s \nonumber \\
\sigma^z&=&- A_l B_l \;. \label{jordanwigner}
\end{eqnarray*}
where $A=c+c^\dagger$ and $B=c-c^\dagger$,
the three-point correlation functions $\langle \sigma^p_i\sigma^q_j\sigma^r_k\rangle$
can be written as a  Pfaffian whose elements are the two-point correlators $\langle A_l B_m\rangle$.
The  structure of the Pfaffians further simplifies in to a determinant
 because  only terms $\langle A_l B_m\rangle=G_{m-l}=
\int^\pi_0\
\frac{1}{\pi}\{\cos(\phi |m-l|)(\cos(\phi)-a)-\gamma\sin(\phi |m-l|)\sin(\phi)\}
\frac{\tanh(\frac{1}{2 T}\sqrt{\gamma^2\sin\phi^2+(h-\cos\phi^2)})}{\sqrt{\gamma^2\sin\phi^2+(h-\cos\phi^2)}}
d\phi$ are non vanishing.

\begin{eqnarray*}
ZZZ_{\alpha\beta}&=&-
\left |
\begin{array}{ccc}
G_{0} & G_{\alpha}&G_{\alpha+\beta}\\
G_{-\alpha} & G_{0}&G_{\beta}\\
G_{-\alpha-\beta} & G_{-\beta}&G_{0}\\
\end{array}\right |,
\end{eqnarray*}
 \begin{eqnarray*}
XXZ_{\alpha\beta}&=&(-)^{\alpha +1} \times \nonumber \\
&&\hskip-20pt \left |
\begin{array}{ccccc}
G_{-1} &G_0 &\ldots & G_{\alpha-2}&G_{\alpha+\beta-1}\\
G_{-2} & G_{-1} & \ldots  & G_{\alpha-3}&G_{\alpha+\beta-2}\\
\vdots & \nonumber & \ddots & \vdots &\vdots \\
G_{-\alpha} & G_{-\alpha-1} & \ldots  & G_{-1}&G_{\beta}\\
G_{-\alpha-\beta} & G_{-\alpha-\beta-1} & \ldots  & G_{-\beta-1}&G_{0}\\
\end{array}\right |,
\end{eqnarray*}
\begin{eqnarray*}
XZX_{\alpha\beta}&=&(-)^{\alpha +\beta} \times \nonumber \\
&&\hskip-20pt \left |
\begin{array}{ccccccc}
G_{-1} &\ldots & G_{\alpha-2}&G_{\alpha}&\ldots & G_{\alpha+\beta-2}\\
\vdots & &\vdots & \vdots  & & \vdots\\
G_{-\alpha+1} &\ldots & G_{0}&G_{2}&\ldots & G_{\beta}\\
G_{-\alpha-1} &\ldots & G_{-2}&G_{0}&\ldots & G_{\beta-2}\\
\vdots & &\vdots & \vdots  & & \vdots\\
G_{-\alpha-\beta} & \ldots &  G_{-\beta-1} &G_{-\beta+1}& \ldots  & G_{-1}\\
\end{array}\right |,
\end{eqnarray*}
 \begin{eqnarray*}
YYZ_{\alpha\beta}&=&(-)^{\alpha +1} \times \nonumber \\
&&\hskip-20pt \left |
\begin{array}{ccccc}
G_{1} &G_2 &\ldots & G_{\alpha}&G_{\alpha+\beta}\\
G_{0} & G_{1} & \ldots  & G_{\alpha-1}&G_{\alpha+\beta-1}\\
\vdots & \nonumber & \ddots & \vdots &\vdots \\
G_{-\alpha+2} & G_{-\alpha+3} & \ldots  & G_{1}&G_{\beta+1}\\
G_{-\alpha-\beta+1} & G_{-\alpha-\beta+2} & \ldots  & G_{-\beta}&G_{0}\\
\end{array}\right |,
\end{eqnarray*}
\begin{eqnarray*}
YZY_{\alpha\beta}&=&(-)^{\alpha +\beta} \times \nonumber \\
&&\hskip-20pt \left |
\begin{array}{ccccccc}
G_{1} &\ldots & G_{\alpha-1}&G_{\alpha+1}&\ldots & G_{\alpha+\beta}\\
\vdots & &\vdots & \vdots  & & \vdots\\
G_{-\alpha+2} &\ldots & G_{0}&G_{2}&\ldots & G_{\beta+1}\\
G_{-\alpha} &\ldots & G_{-2}&G_{0}&\ldots & G_{\beta-1}\\
\vdots & &\vdots & \vdots  & & \vdots\\
G_{-\alpha-\beta+2} & \ldots &  G_{-\beta} &G_{-\beta+2}& \ldots  & G_{1}\\
\end{array}\right |.
\end{eqnarray*}

The main difference with respect the two-point correlators calculated in \cite{MODELLO}
is that in this case the Toeplitz structure of the matrix is no longer valid.

\end{document}